\documentclass[11pt]{amsart}
\usepackage{amsmath,amsthm,amscd,amsfonts,amssymb}
\usepackage[all]{xy}
\usepackage[ansinew]{inputenc}

\usepackage{hyperref}

\newtheorem{thm}{Theorem}[section]

\theoremstyle{definition}
\newtheorem{defi}[thm]{Definition}

\newtheorem*{lemacero}{Lema 0}
\theoremstyle{remark}

\numberwithin{equation}{section}

\theoremstyle{remark}

\newcommand{\cC}{{\mathcal C}}

\begin{document}

\title[A note on the foundations of mechanics]
 {A note on the foundations of mechanics}

\author{R. J.  Alonso-Blanco and J.  Mu\~{n}oz-D{\'\i}az}

\address{Departamento de Matem\'{a}ticas,
  Universidad de Salamanca, Plaza de la Merced 1-4, E-37008 Salamanca,
  Spain.}
\email{ricardo@usal.es, clint@usal.es}

\begin{abstract}
This short note is intended to review the foundations of
mechanics, trying to present them with the greatest mathematical
and conceptual clarity. It was attempted to remove most of
inessential, even parasitic issues, which can hide the true nature
of basic principles. The pursuit of that goal results in an
improved understanding of some topics such as constrained systems,
the nature of time or the relativistic forces. The Srödinger and
Klein-Gordon equations appear as conditions fulfilled by certain
types of classical solutions of the field equations although the
meaning of quantum equations is not, even remotely, exhausted by
these cases. A part of this note comes from previous works
\cite{RAlonso, MecanicaMunoz, RelatividadMunoz}.
\end{abstract}

\maketitle

\centerline{\today}

\vskip 1cm

\setcounter{section}{-1}
\section{Notation and review on the principles of Lagrangian mechanics}

Let $M$ be a smooth manifold, $TM$ its tangent bundle. A vector
tangent to $TM$ or a vector field tangent to $TM$ are said to be
\emph{ vertical} when annihilate the subring $\cC^\infty(M)$ of
$\cC^\infty(TM)$. An 1-form on $TM$ is said to be \emph{
horizontal} when it is incident with all of the vertical vectors.
In local coordinates $(x^1,\dots,x^n,\dot x^1,\dots,\dot x^n)$ on
$TM$, the vertical vectors or fields are combinations of
$\partial/\partial\dot x^i$, and the horizontal forms are of
$dx^i$. A horizontal 1-form $\alpha$ defines a function
$\dot\alpha$ on $TM$, by the rule $\dot\alpha(v_a)=\langle
\alpha,v_a\rangle$, for each $v_a\in TM$. In particular, for each
function $f\in\cC^\infty(M)$, $\dot{\overline{df}}$ is denoted
simply by $\dot f$. The application
$$\dot d\colon\cC^\infty(M)\to\cC^\infty(TM),\quad f\mapsto \dot
d(f)=\dot f$$ is, essentially, the differential.

A tangent field $D$ on $TM$ such that, as a derivation
$\cC^\infty(M)\to\cC^\infty(TM)$, coincides with $\dot d$ is a
\emph{second order differential equation}; for each $v_a\in T_aM$
is, then, $\pi_*(D_{v_a})=v_a$.

Since two second order differential equations act identically on
$\cC^\infty(M)$, they differ by a vertical field. For this reason,
second order differential equations are the sections of an affine
fiber bundle on $TM$, modeled on the vector bundle of vertical
fields. From now on, a vertical field will be called \emph{ a
force}; a vertical tangent vector $V_{w_a}\in T_{w_a}TM$, will be
called \emph{a force at} $w_a$. The value of a second order
differential equation at a point $w_a\in TM$ will be called
\emph{an acceleration} (at the velocity $w_a$).

The vector structure of the fibers of $TM$, identifies canonically
each force $V_{w_a}\in T_{w_a}T_aM$ with a vector $v_a$ of the
fiber $T_aM$; this vector $v_a\in T_aM$ will be called the
\emph{geometric representative of the force} $V_{w_a}$.

Let $T^*M$ be the cotangent bundle to $M$. The \emph{Liouville
form} $\theta$ on $T^*M$ is defined by
$\theta_{\alpha_a}=\alpha_a$ at each $\alpha_a\in T^*M$. Its
exterior differential $\omega_2=d\theta$ is the \emph{symplectic
form} on $T^*M$. In local coordinates
$(x^1,\dots,x^n,p_1,\dots,p_n)$ of $T^*M$ we have
$\theta=p_1\,dx^i$, $\omega_2=d\theta=dp_i\wedge dx^i$.

Let $T_2$ be a pseudoriemannian (= non degenerate) metric on $M$;
$T_2$ establishes an isomorphism $TM\simeq T^*M$, by assigning to
each vector $v_a\in TM$ the 1-form $\alpha_a=i_{v_a}T_2$; we will
say that $v_a$ is the \emph{gradient} of $\alpha_a$.  When we let
the vector $v_a$ to vary along $TM$ we obtain $i_{\dot
d}T_2=\theta$: the tautological field $\dot d$ (which is the
identity $TM\to TM$) corresponds with the tautological form
$\theta$.

The isomorphism $TM\simeq T^*M$ produced by the metric $T_2$ allow
us to transport the structures from each bundle to the other. For
this reason, we talk about the Liouville and the symplectic forms
on $TM$, by using the same notation. The function $T=\frac
12\,\dot\theta$ is the \emph{kinetic energy} of $(M,T_2)$.
\bigskip

The whole of the Mechanics rests on the following
\begin{lemacero}
The metric $T_2$ establishes a biunivocal correspondence between
second order differential equations $D$ and horizontal 1-forms
$\alpha$ by means of the equation
\begin{equation}\label{lema0}
i_D\omega_2+dT+\alpha=0.
\end{equation}
The tangent fields $u$ on $M$ which are intermediate integrals of
$D$ are those holding
\begin{equation}\label{lema01}
i_u\,d(i_uT_2)+dT(u)+u^*\alpha=0,
\end{equation}
where $u^*\alpha$ is th pull-back of $\alpha$ by means of $u\colon
M\to TM$ and $T(u)$ is the value of the function $T$ when
specialized to $u$.
\end{lemacero}

A \emph{classical mechanical system} is a manifold $M$ (the
\emph{space of configuration}) endowed with a pseudoriemannian
metric $T_2$ and a horizontal form $\alpha$, the \emph{work form}.
The second order differential equation $D$ which corresponds with
$\alpha$ according (\ref{lema0}) is the \emph{differential
equation of the motion}  and (\ref{lema0}) is the \emph{Newton
equation} of the system $(M,T_2,\alpha)$.

The tangent field which corresponds with $\alpha=0$ is called
\emph{geodesic field}; it will be denoted by $D_G$. For an
arbitrary system $(M,T_2,\alpha)$, the difference $D-D_G$ is a
vertical field, the \emph{force} of the system. The geometric
representative of the force is a field on $TM$ with values in
$TM$, which will be denoted by $D^\nabla$ and called
\emph{covariant value} of $D$; it holds
$i_{D^\nabla}T_2+\alpha=0$, or, what is the same
$D^\nabla=-\textrm{grad}\,\alpha$, and that is the general form of
the Newton's law `\emph{force=mass$\cdot$acceleration}', by
observing that, on each curve solution of $D$ on $M$ it holds
$D^\nabla=u^\nabla u$, when $u$ is the tangent vector along such a
curve.

A system of \emph{constraints} on $M$ is a Pfaff system $\Lambda$
on $TM$ comprised by horizontal forms. The Newton equation
(\ref{lema0}) for a system with constraints
$(M,T_2,\alpha,\Lambda)$ is substituted by a congruence
\begin{equation}\tag{\ref{lema0}'}\label{Newtonligada}
i_D\omega_2+dT+\alpha\equiv 0\,\, (\textrm{mod}\,\Lambda)
\end{equation}
joint with the \emph{Virtual Works Principle}: the trajectories of
the system remain into the subset of $TM$ defined by
$\dot\beta=0$, $\forall\beta\in\Lambda$.

The determination of the field $D$ by condition
(\ref{Newtonligada}) plus the virtual works principle is possible
under suitable conditions on the metric and the constraints; for
example, when the metric is positive definite and $\Lambda$ is the
extension to $TM$ of a regular Pfaff system on $M$. We will not
dwell further on this issue, which now is not of direct interest
for our purposes.

\section{Conservative systems}

When $\alpha$ is an exact differential form, the system
$(M,T_2,\alpha)$ is said to be \emph{conservative}; since $\alpha$
is horizontal, the potential function $V$ of which $\alpha$ is its
differential, must belong to the subring $\cC^\infty(M)$ of
$\cC^\infty(TM)$. The sum $H=T+V$ is called the \emph{hamiltonian}
of the system and the Newton equation (\ref{lema0}) is, in this
case,
\begin{equation}\label{newtonconservada}
i_D\omega_2+dH=0.
\end{equation}
When written in coordinates $(x^i,p_i)$ of $T^*M$,
(\ref{newtonconservada}) is the system of  \emph{Hamilton
canonical equations}. From (\ref{newtonconservada}) and the
classical argument based on Stokes theorem, it follows the
\emph{Maupertuis principle}: the trajectories of the mechanical
system on $M$ are extremal for $\int\theta$ with fixed end points
and $H=\textrm{const.}$

The equation (\ref{lema01}) for the intermediate integrals becomes
in this case:
\begin{equation}\label{intermediaconservada}
i_ud(i_uT_2)+dH(u)=0
\end{equation}
In particular, the fields $u$ which, as submanifolds of $TM$, are
lagrangian, hold $d(i_uT_2)=d\theta\mid_u=0$, so that
\begin{equation}\label{intermedialagrangiana}
dH(u)=0\quad\text{or}\quad H(u)=\textrm{const.}
\end{equation}
This is the \emph{Hamilton-Jacobi equation}.

Let us demand somewhat more of $u$: to be, in addition to
lagrangian, conservative, that is to say, of null divergence.
Since $u$ is lagrangian, locally  we have $u=\textrm{grad}\,f$,
for a suitable function $f$ in $M$; then, $\textrm{div}\,u=\Delta
f=0$.

We have a formula, which holds for $T_2$ without conditions,

\begin{equation}\label{laplaciana}
\Delta
e^{i\phi}=e^{i\phi}\left(i\Delta\phi-T^2(d\phi,d\phi)\right),
\end{equation}
for all local function $\phi$ on $M$. By applying it in our case
to the harmonic function $\phi=f/h$ ($h$ is a constant) and
calling $\Psi:=e^{if/h}$ we get
 $$
 \Delta\Psi =-\frac\Psi{h^2}\,T^2(df,df)
 =-\frac\Psi{h^2}\,T_2(u,u)
 =-\frac {2\,\Psi}{h^2}(H(u)-V),
 $$
 and then, if $E$ is the constant value of $H(u)$:
 \begin{equation}\label{Sch}
 \left(\frac {h^2}2\Delta-V\right)\Psi=-E\Psi
 \end{equation}
 This is the \emph{Schrödinger equation}. Conversely, if (\ref{Sch}) holds
  $\parallel\Psi\parallel=1$ ($E$ real), then we get an intermediate integral which is
  lagrangian and conservative. In addition, by taking the real or the imaginary parts of these type of
  fields $\Psi$, we get solutions which are not interpretable in classical terms.

 \section{Lenght and time}

 In Mechanics, \emph{time} is the parameter of each
 curve-solution of field $D$ which govern the evolution of the
 system. However, it cannot be a function on $TM$ which serves as
 a parameter for all the solutions of all the second order
 differential equations on $M$; on $TM$ does not exist any
 function which can be called time. The natural object which
 serves to parametrice all the curve-solutions of all the second
 order differential equations is a class of 1-forms, the
 \emph{class of time}, comprised by the local horizontal 1-forms
 $\alpha$ such that $\dot\alpha=1$; for each horizontal 1-form
 $\beta$, in the open subset where $\dot\beta\ne 0$, the quotient
 $\beta/\dot\beta$ belongs to the class of time. Two forms
 belonging to the class of time differ, in their common domain, by
 a form of the contact system $\Omega$ of $TM$; the class of time
 is that of $\theta/\dot\theta\,\textrm{mod}\,\Omega$.

 We can \emph{choose a time} for $(M,T_2)$ by giving a horizontal
 form $\tau$ on $TM$ and considering as possible trajectories just
 those which are contained in the subset $\dot\tau=1$ of $TM$; with
 this, the space of states is restricted to that subset. Assuming
 that $\dot\tau=1$ or, more generally, $\dot\tau=\textrm{const.}$
 defines a submanifold of $TM$, for each mechanical system
 $(M,T_2,\alpha)$, the field $D$ governing the evolution prior to
 the time constraint, must be modified to another field $\overline
 D$ holding the congruence
 \begin{equation}\tag{\ref{lema0}''}\label{Newtontiempo}
i_{\overline D}\omega_2+dT+\alpha\equiv 0\,\, (\textrm{mod}\,\tau)
\end{equation}
and, in substitution of the Virtual Works Principle, the
condition: $\overline D$ is tangent to the manifolds
$\dot\tau=\textrm{const.}$

For instance, when $\tau$ is a form on $M$, we have
\begin{equation}\label{Campotiempo}
\overline D=D-\frac 1{\parallel \tau\parallel^2}
  \left(\langle\tau,D^\nabla\rangle+\textrm{II}_{\textrm{grad}\,\tau}(\dot
  d,\dot d)\right)\,\textrm{Grad}\,\tau
\end{equation}
out of the set where $\parallel\tau\parallel=0$; that set is void
in the most classical case: $\tau$ without zeroes and $T_2$
positive definite. In (\ref{Campotiempo}), $\textrm{Grad}\,\tau$
is the vertical field which corresponds to $\textrm{grad}\,\tau$
and $\textrm{II}_u$ is the second fundamental form of the tangent
field $u$ with respect to $T_2$.

The form
$\frac{\theta}{\parallel\theta\parallel}=\frac{\theta}{\sqrt{\mid\dot\theta\mid}}$
is defined on the open set of $TM$ where $\dot\theta (=2T)\ne 0$;
when $T_2$ is positive definite, that open is $TM$ without the
zero section; when $T_2$ has the signature of Minkowski, the open
set excludes the `light cones'', etc.

In its domain $\frac{\theta}{\parallel\theta\parallel}$ is
invariant under the group of homoteties on fibers; in fact, under
the whole of groups generated by fields $\mu V$, where $V$ is the
infinitesimal generator of homoteties and $\mu\in\cC^\infty(TM)$.
For this reason, $\frac{\theta}{\parallel\theta\parallel}$
projects to the open of the space of 1-jets of curves $J_1^1M$
which is image of $\dot\theta\ne 0$.

Physicists call the length \emph{proper time}; this is why we can
call $\frac{\theta}{\sqrt{\mid\dot\theta\mid}}$ the \emph{lenght
form} or the \emph{proper time form}.

A non-parametrized curve $\gamma$ in $M$, defines canonically a
curve $\widetilde\gamma$ in $J_1^1M$; the integral of
$\frac{\theta}{\parallel\theta\parallel}$
 along $\widetilde\gamma$ is the \emph{length} of $\gamma$,
 which is independent of any parametrization: it is the fact of $\frac{\theta}{\parallel\theta\parallel}$
being projectable by $TM\to J_1^1M$ what enables us to define the
length or proper time. Except for some technical detail, similar
reasons allow us to define the volume of r-dimensional submanfolds
of $M$ ($1\le r\le n$).

\section{Relativistic forces}

We have considered the forms $\theta$,
$\frac{\theta}{\dot\theta}$,
$\frac{\theta}{\parallel\theta\parallel}$, the last two defined on
the open $\dot\theta\ne 0$. The form $\frac{\theta}{\dot\theta}$
and all those which are congruent with it $\textrm{mod}\,\Omega$,
specialize on each curve solution of each second order
differential equation $D$, as the differential of the natural
parameter, let us say $dt$. The form
$\frac{\theta}{\parallel\theta\parallel}=\frac{\theta}{2T}$
specializes on each curve of $TM$ as the length form or length
element $ds$; its integral along such a curve is the length of the
curve once projected to $M$. Physicists parametrize the
``relativistic'' motions of particles by the ``proper time'' of
each particle. Since in a mechanical system, without any condition
on the configuration space $M$ or the metric $T_2$, the unique
possible parameter for the trajectories (of second order
differential equations) is the class of time, it follows that on
the relativistic trajectories must be suitable as a natural
parameter, both $\frac{\theta}{\dot\theta}$ and
$\frac{\theta}{\parallel\theta\parallel}$. That condition
determines uniquely if a trajectory is relativistic or not: a
curve solution of a second order differential equation on $TM$ is
``relativistic'' if on it holds $\dot\theta=1$, $\dot\theta=-1$ or
$\dot\theta=0$.

This is why, provisionally, it is natural to distinguish between
relativistic or no relativistic mechanical systems according o the
following criterium: $(M,T_2,\alpha)$ is relativistic if the field
$D$ which governs the evolution of such a system is tangent to the
hypersurfaces $\dot\theta=1$, $\dot\theta=-1$, $\dot\theta=0$ of
$TM$.

Let us decompose the field $D$ as a sum $D=D_G+W$, where $D_G$ is
the geodesic field and $W$ is a vertical field. $D_G$ is
relativistic; its Newton equation is $i_{D_G}\omega_2+dT=0$, and
then $D_G\dot\theta=D_G(2T)=0$. Hence, $D$ is tangent to the
hypersurfaces $\dot\theta=0,\pm 1$, if and only if $W$ it is. Now,
since $W$ is vertical, we have $W\dot\theta=\langle
w,\theta\rangle$, where $w$ is the geometric representative of $W$
($w=D^\nabla$). In local coordinates, $\langle
w,\theta\rangle=g_{ij}w^i\dot x^j$; if the functions $w^i$ are
homogeneous (of arbitrary degree) on the $\dot x$, $\langle
w,\theta\rangle$ is homogeneous in the $\dot x$, hence, if
vanishes on $\dot\theta=1$ then vanishes on each
$\dot\theta=\textrm{const.}$

As a consequence, the relativistic fields $D=D_G+W$ when $W$ is an
homogeneous of the velocities $\dot x$, are tangent to all of the
submanifolds $\dot\theta=\textrm{const}$; this is to say,
$\dot\theta$ must be a first integral of $D$, hence, also of $W$;
from the equation (\ref{lema0}) follows $DT+\dot\alpha=0$, and
then $DT=0$ $\Leftrightarrow$ $\dot\alpha=0$ $\Leftrightarrow$
$WT=0$.

Summing up: it is natural to adopt the following
\begin{defi}
The mechanical system $(M,T_2,\alpha)$ is relativistic if the
corresponding field $D$ holds $D\dot\theta=0$. This condition is
equivalent to the field of forces $W=D-D_G$ holding
$W\dot\theta=0$.
\end{defi}
And it holds the
\begin{thm}
$(M,T_2,\alpha)$ is relativistic if and only if the work form
$\alpha$ belongs to the contact system $\Omega$ of $TM$.
\end{thm}

This condition is independent of the metric!

Since the contact system does not contain exact forms, except the
0 form, there is no other relativistic and conservative system but
the geodesic one. The fact that the conservative systems (except
the geodesic) never are relativistic, has nothing to do with the
absence of action at a distance o the finiteness of the speed of
light, because it is an exclusive consequence of the lack of
closed forms in the contact system.

In local coordinates, a system of generators of $\Omega$ is the
comprised by forms $\dot x^h\,dx^j-\dot x^j\,dx^h=i_{\dot
d}(dx^h\wedge dx^j)$. It follows that for every 1-form $\alpha$ of
the contact system $\Omega$ there exists a horizontal
hemisymmetric tensor field (a horizontal 2-form) $F_2$ on $TM$
such that $i_{\dot d} F_2=\alpha$: every relativistic force is
``produced'' by an hemisymmetric covariant tensor field, of order
2,  horizontal on $TM$; in general, the field $F_2$ is not
completely determined by $\alpha$. When, the force field $W$
depends linearly on the velocities (that is to say, of the $\dot
x$'s), also is this way for $\alpha=-i_wT_2$; in such a case,
there exists a unique 2-form $F_2$ on $M$ such that $i_{\dot
d}F_2=\alpha$; the 2-forms on $M$ are the same objects as the
contact 1-forms on $TM$ which linearly depend on the $\dot x$. We
have
\begin{thm}
The relativistic forces which depend linearly on the velocities
correspond biunivocally with 2-forms on $M$. To the 2-form $F_2$
it corresponds the force 1-form $\alpha=i_{\dot d}F_2$.
\end{thm}

\section{Electromagnetic fields}

By an extension of the classical case, we will call
\emph{electromagnetic field} on $M$ to each closed 2-form $F_2$ on
$M$. The 1-form $\alpha=i_{\dot d}F_2$, which is the translation
of $F_2$, will be called \emph{Lorentz force form} of the
electromagnetic field. This form does not depend on the metric on
$M$; once the metric is fixed, the Newton equation
\begin{equation}\tag{\ref{lema0}'''}\label{NewtonLorentz}
i_D\omega_2+dT+i_{\dot d}F_2=0
\end{equation}
determines  the second order differential equation $D$, to which
we can call the \emph{Lorentz field} produced by the
electromagnetic field $F_2$ on $(M,T_2)$; the associated force
$D-D_G$ is the \emph{Lorentz force}.

Looking at the Newton equation (\ref{NewtonLorentz}), it is
natural to consider the following 2-form on $TM$:
\begin{equation}\label{Nuevaomega}
\omega_F=\omega_2+F_2.
\end{equation}

$\omega_F$ is a new symplectic form on $TM$; the Newton equation
(\ref{NewtonLorentz}) is written:
\begin{equation}\label{NewtonLorentznueva}
i_D\omega_F+dT=0
\end{equation}

Thus, the Lorentz field $D$ is the hamiltonian field which
corresponds to the hamiltonian $T$ in the symplectic structure
$\omega_F$.

The equation (\ref{lema01}) is, for the intermediate integrals of
the Lorentz field:
\begin{equation}\label{intermediaLorentz}
i_u(F_2+di_uT_2)+dT(u)=0
\end{equation}
or
\begin{equation}\tag{\ref{intermediaLorentz}'}\label{intermediaLorentz2}
i_u({\omega_F}\mid_u)+dT(u)=0
\end{equation}
where ${\omega_F}\mid_u$ is the specialization of $\omega_F$ to
the section $u$ of $TM$.

The intermediate integrals of the Lorentz field which are
lagrangian for the symplectic form $\omega_F$ hold
\begin{equation}\tag{\ref{intermediaLorentz}'}\label{intermediaLorentz3}
{\omega_F}\mid_u=0
\end{equation}
hence
\begin{equation}\label{intermediaLorentz4}
T(u)=\textrm{const.}
\end{equation}
which is the Hamilton-Jacobi equation for the
$\omega_F$-lagrangian intermediate integrals of $D$.

The closed 2-form $F_2$ is locally exact; locally $F_2$ can be
written as $F_2=d(i_AT_2)$, where the field $A$ is the
\emph{potential vector} for $F_2$. The fact that a tangent field
$u$ on $M$ is a section a $\omega_F$-lagrangian section of $TM$ is
written as $$d(i_{u+A}T_2)=0,$$ so that
\begin{equation}\label{potencialLorenz}
i_{u+A}T_2=df
\end{equation}
locally, with $f\in\cC^\infty(M)$.

If the field $u+A$ is conservative ($\textrm{div}(u+A)=0$), we
have $\delta df=\Delta f=0$ and the formula (\ref{laplaciana})
applied on $\Psi=e^{i\frac fh}$ ( $h$ is a constant) gives us
\begin{align*}
\Delta\Psi&=-\frac\Psi{h^2}\,T^2(df,df)\\
&=-\frac\Psi{h^2}\,
T_2(u+A,u+A)\\
&=-\frac\Psi{h^2}\left(\parallel u\parallel^2+\parallel
A\parallel^2+2\, T_2(A,u)\right)\\
&= -\frac\Psi{h^2}\left(\parallel u\parallel^2+\parallel
A\parallel^2+2\ \langle A,df\rangle-2\,T_2(A,A)
\right)\\
&=-\frac\Psi{h^2}\left(\parallel u\parallel^2-\parallel
A\parallel^2+2\, A(f)\right)\\
&=-\frac 1{h^2}\left(\parallel u\parallel^2-\parallel
A\parallel^2-2\,ihA\right)\Psi.
\end{align*}
Equation (\ref{intermediaLorentz4}) gives us a constant $\parallel
u\parallel^2=m^2$; then, it follows
\begin{equation}\label{KG}
\left(\Delta-2\,\frac ih\,A+\frac 1{h^2}\,(m^2-\parallel
A\parallel^2)\right)\Psi=0
\end{equation}
which is the Klein-Gordon equation:
\begin{thm}
The Klein-Gordon equation (\ref{KG}) characterizes the tangent
fields $u$ on $M$ which are intermediate integral of the Lorentz
field $D$, lagrangian with respect the symplectic structure
$\omega_F=\omega_2+F_2$ and such that $\textrm{div}\,(u+A)=0$,
where $A$ is a vector potential for $F_2$.
\end{thm}

\noindent\textbf{Remark on th first pair of Maxwell equations} The
first pair of Maxwell equations $\delta F_2=J^\star$ is
interpretable as a \emph{definition} of electric current. However,
it is not true, in the classical case, that the ``electric fluid''
has necessarily as pathlines the curves solution of the field $J$,
because we can change $F_2$, and with it the Lorentz force,
without any change of $J$. In the classical case,
$M=\mathbb{R}^4$-Minkowski, we can choose the vector-potential $A$
in such a way that it holds the ``Lorentz gauge'' condition,
$\textrm{div}\,A=0$ and, then, (\ref{KG}) is a condition on the
intermediate integral $u$: $u$ is lagrangian and
$\textrm{div}\,u=0$. If we assume that $J$ is a particle flow
which obey to the Lorentz force, $J=u$ is an intermediate integral
of that force and automatically, by its very definition
$J^\star=\delta F_2$, it holds $\textrm{div}\,J=0$ and the
Klein-Gordon equation (\ref{KG}) is the condition to the current
$J$ be lagrangian.

\end{document}